\documentclass[aps,prl,twocolumn,showpacs,letterpaper]{revtex4}
\usepackage{amsmath, amsthm, amssymb}
\usepackage{graphicx}

\newcommand{\diss}{ \ensuremath{\mathcal D}}
\newcommand{\JS}{\mathrm{JS}}

\begin{document}
\title{Far-from-Equilibrium Measurements of Thermodynamic Length}
\date{\today}
\author{Edward H.\ Feng}
\affiliation{College of Chemistry, University of California, Berkeley, Berkeley,
California 94720, USA}
\author{Gavin E.\ Crooks}
\affiliation{Physical Biosciences Division, Lawrence Berkeley National Laboratory, Berkeley,
California 94720, USA}

\begin{abstract}
Thermodynamic length is a path function that generalizes the notion of length to the surface of  thermodynamic states. 
Here, we show how to measure thermodynamic length in far-from-equilibrium single molecule experiments using the work fluctuation relations.
For these microscopic systems, it proves necessary to define the thermodynamic length in terms of the Fisher information. 
 Consequently, the thermodynamic length  can be directly related to the magnitude of fluctuations about equilibrium. 
The work fluctuation relations link the work and the free energy change during an external perturbation on a system. 
We use this result to determine equilibrium averages at intermediate points of the protocol in which the system is out-of-equilibrium. 
This allows us to extend Bennett's method to determine the potential of mean force, as well as the thermodynamic length, in single molecule experiments. 
\end{abstract}
\preprint{}
\maketitle


Modern experimental techniques
allow the manipulation of single
molecules and the measurement of the thermodynamic properties of 
microscopic systems~\cite{Liphardt2002,Bustamante2005,Collin2005,Ritort2006a,Maragakis2008b}. 
For example, Collin et al.~\cite{Collin2005}  recently measured the work performed on a single RNA hairpin as it was folded and unfolded using optical tweezers. From these out-of-equilibrium measurements, they extracted
 the equilibrium free energy change using the recently discovered work
fluctuation relations~\cite{Jarzynski1997a,Crooks1999a,Hummer2001a}.~[Eq.~({\ref{eq:WFT})]
These relations, which connect the free energy change and the work done on a system by an external perturbation, remain valid no matter how far the system is driven away from thermal equilibrium.

In this Letter, we will demonstrate that free energy is not the only important quantity that can be extracted from out-of-equilibrium work measurements; we can also measure the thermodynamic length~\cite{Weinhold1975a,Weinhold1976,Ruppeiner1979,Salamon1983a,Salamon1985,Nulton1985b,Schlogl1985,Crooks2007c}.  Thermodynamic length is a path function that measures the distance along a path in thermodynamic state space. This is in contrast to the free energy change, a state function which depends only on the initial and final values of the controllable parameters, and not on the path.  Mathematically, the thermodynamic length is defined by a Riemannian metric on the manifold of equilibrium ensembles~\cite{Burbea1982,Brody1995}. Among other useful physical properties, the  thermodynamic length bounds the dissipation of slow, but finite time transformations~\cite{Salamon1983a, Nulton1985b}. Moreover, the ability to measure thermodynamic length and free energy change from out-of-equilibrium measurements  indicates that these equilibrium properties influence the behavior of driven systems even far-from-equilibrium.

Thermodynamic length was originally defined using the
second derivatives of a thermodynamic potential with respect to 
its natural  variables~\cite{Weinhold1975a,Weinhold1976}.   
However, this definition only works for microscopic systems when the controlled variables are intensive
(e.g.\ temperature)~\cite{Crooks2007c}.
To circumvent this restriction, herein we will  
redefine the thermodynamic length in terms of Fisher information~\cite{Rao1945,Burbea1982}. This approach is equivalent to the original definition for large systems in the  thermodynamic limit~\cite{Brody1995,Crooks2007c}, but can also be applied, without restriction, to microscopic systems, or to problems outside of thermodynamics entirely. 

Given a family of probability distributions $\pi(x|\lambda)$ for outcomes $x$ that vary smoothly with a collection of parameters $\lambda= \{ \lambda^i \}$, the Fisher information matrix~\cite{Fisher1925,Cover1991} is 
\begin{equation}
{\mathcal I} _{ ij } ( \lambda ) \equiv \int dx\
\pi (x|\lambda )\ \frac{\partial \ln \pi (x| \lambda) }{
\partial \lambda ^{i} } \ \frac{\partial \ln \pi (x| \lambda) }{
\partial \lambda ^{j} } 
\ .
\end{equation}
%
The length of a path $\lambda (s) $ for $s\in [0,1] $ in parameter space measured using the Fisher metric (also known as the Fisher-Rao, Rao or entropy differential metric) is~\cite{Rao1945}
\begin{equation}
\mathcal{L} = \int _{0} ^{ 1}  \left[
\sum_{ij}
\frac{ d \lambda ^{i}(s)   }{ds }\  {\mathcal I} _{ ij }(\lambda(s))\  \frac{d \lambda ^{j}(s)  }{ds }
 \right] ^{ 1/2 }  ds . 
\label{eq:raolength}
\end{equation}
The Fisher matrix ${\mathcal I}_{ij}$ acts as a metric tensor and equips the manifold of parameters with a Riemannian metric~\cite{Rao1945,Burbea1982}.
It is also useful to define a related quantity, the Fisher divergence 
\begin{equation}
\mathcal{J} \equiv \int _{0} ^{ 1}  
  \sum_{ij} \frac{ d \lambda ^{i}(s)  }{ds } \ {\mathcal I} _{ ij }(\lambda(s))\  \frac{d \lambda ^{j}(s)  }{ds } 
\   ds  \ .
\label{eq:divergence}
\end{equation}
The length and divergence are connected by the relation
${\mathcal J } \geq {\mathcal L}^2$ due to the Cauchy-Schwarz inequality.
 
The Fisher metric can be applied to any family of probability distributions.
Here, we focus on probability distributions of a system in thermal
equilibrium.  
In the canonical ensemble~\cite{Gibbs1902,Chandler1987a}, the probability of 
a micro-state $x$ is 
\begin{align}
\pi(x |\lambda)  = \exp\Big\{ \beta F(\lambda) - \beta E(x,\lambda)\Big\} \ ,
\label{eq:canonical}
\end{align}
where $\beta= 1/k_{\text{B}} T$ is the inverse temperature $T$ of the environment in natural units ($k_{\text{B}}$ is the Boltzmann constant), $E(x,\lambda)$ is the energy of the system, which depends both on the internal state $x$ and the external control parameters $\lambda$,  and $F(\lambda)$ is the free energy
\begin{equation}
 \beta F (\lambda) = -\ln \sum_x  \exp\{-\beta E(x,\lambda) \} \ .
\end{equation}

For a case in which $\lambda $ is a single controllable parameter, 
 the Fisher information is
 \begin{equation}
 I(\lambda)= \beta^2 \left \langle\left( \frac{d F (\lambda)}{d \lambda} -  \frac{\partial E(x,\lambda )}{\partial \lambda}   \right)^2 \right\rangle _{ \lambda } 
 \ ,
 \end{equation}
 where $\langle\cdots\rangle _{ \lambda } $ indicates an ensemble average 
over the distribution $\pi (x| \lambda )  $.

Let us consider two examples. First, suppose the system under examination is a single polymer, and the parameter under control is the end-to-end distance $L$ (Concretely, an RNA hairpin, with DNA handles, attached to beads held by a translating optical trap~\cite{Collin2005}). The instantaneous tension $\mathcal{T} = \frac{\partial E}{\partial L}$ is the force exerted on the polymer by the apparatus constraining the distance between the polymer ends. The Fisher information for this system is equal to the variance of the tension 
at equilibrium
\begin{align}
\mathcal{I}(L) 
 &= \beta^2 \left \langle\big( \left\langle {\mathcal T}  \right\rangle -  {\mathcal T}    \big)^2
  \right\rangle 
  \ ,
\end{align} 
and therefore the thermodynamic length  [Eq.~(\ref{eq:raolength})] is equal to the cumulative root-mean-square equilibrium fluctuations in tension on the molecule.

On the other hand, suppose control is exerted by applying constant tension to the ends of the polymer. The total energy is then a linear function of length and tension, $E(x,  {\mathcal T} ) = U(x) -  {\mathcal T} L(x)$, and the Fisher information is equal to the variation of the end-to-end polymer length at equilibrium.
\begin{align}
\mathcal{I}({\mathcal T} ) 
 &= \beta^2 \left \langle\big( \left\langle L  \right\rangle -  L    \big)^2
  \right\rangle 
\end{align}
Again, the Fisher information  has a simple physical interpretation
in terms of equilibrium fluctuations, and the thermodynamic length [Eq.~(\ref{eq:raolength})] is equal to the cumulative root-mean-square fluctuations along the path. 
If, as in the second case, the energy is a linear function of the control parameter, the Fisher information is equal to the second derivative of the free entropy~\cite{Schlogl1985, Brody1995, Crooks2007c}, but this is not true in the general.

To demonstrate how to measure thermodynamic length in far-from-experiments, we will model the dynamics of the system as a driven, discrete time,  inhomogeneous
Markov process~\cite{Crooks1998,Crooks1999a,Crooks2000}. The microscopic state of the system is denoted by $x$, and the history of the system will be denoted by $x _{a,b }   \equiv \{ x _{a} , x _{a+1} , \cdots, x _{ b } \} $, where $a\leq b$. The time reversed trajectory is  denoted by 
 $\tilde{x} _{ b,a }\equiv \{ x _{ b }, x _{ b-1 },\cdots , x _{a} \} $.  The internal energy $E(x,\lambda)$ depends on a  control parameter $\lambda$ that  varies according to a predetermined protocol, $\Lambda_{a,b} \equiv \{\lambda_a, \lambda_{a+1}, \cdots, \lambda_{b} \}$. The protocol of the conjugate  time reversed experiment is $\tilde{\Lambda}_{b,a} \equiv \{\lambda_{b}, \lambda_{b-1}, \cdots, \lambda_{a} \}$. At each integer time $t$, the control value  changes from $\lambda_t$ to $\lambda_{t+1}$ in the forward protocol and $\lambda_{t+1}$ to $\lambda_{t}$ in the reversed protocol. Between these time points $\lambda$ is constant.

 The probability of observing a particular trajectory $x_{a,b}$ as the system is driven from thermal equilibrium by the protocol $\Lambda_{a,b}$ can be written as
 \begin{equation}
P _{ a, b} [ x _{ a,b } ] \equiv \pi( x _{a}| \lambda_a ) \prod_{t= a}^{b-1} p _{t} ( x _{ t+1 },x _{t} ) 
\label{eq:pathprobforward}
\end{equation}
where $\pi( x _{a}| \lambda_a )$ is the initial equilibrium probability distribution with fixed $\lambda$ [Eq.~({\ref{eq:canonical})], and $p _{t} ( x _{ t+1 },x _{t})$ is the probability of transitioning from state $x_{t}$ at time $t$ to state $x_{t+1}$ at $t+1$, given the value of $\lambda$ at time $t$.
These transition probabilities satisfy the stochastic property
\begin{equation}
\sum_{ x' } p _{t} ( x', x) = 1
\label{eq:stochastic}
\end{equation}
which ensures conservation of probability, and the balance condition
\begin{equation}
\sum_{ x } p _{t} ( x', x) \pi(x|\lambda_t) = \pi(x'|\lambda_t)
\end{equation}
which ensures that the canonical ensemble is the stationary distribution of the dynamics. 

The transition probabilities of the  forward and time reversed dynamics are related, since, at equilibrium, the transition $x \rightarrow x' $ in the forward dynamics has the same probability as the transition $x' \rightarrow x$  in the reversed dynamics~\cite{Norris1997,Crooks2000}, given a fixed $\lambda$.
 Explicitly, the reverse time transitions are related to the forward time transitions by~
 \begin{equation}
\tilde{p} _{ t } ( x , x' ) \ 
\pi (x' |\lambda_t)
= 
p _{t} ( x',x )\  \pi( x |\lambda_t) \,.
\label{eq:reverse}
\end{equation}
A direct consequence of this time reversal symmetry is 
the work fluctuation theorem~\cite{Crooks1999a,Crooks2000,Collin2005}: 
the ratio of the probabilities of the forward and reverse trajectories is the  exponential of the observed dissipation along the forward trajectory.
\begin{align}
\frac{P _{ a,b } [x _{ a,b } ] }{
P _{ b,a } [ \tilde{x} _{ b,a } ]}
& = e ^{ \beta W _{ a,b } [x _{ a,b } ] - \beta \Delta F _{ a,b } } 
\label{eq:WFT}
\\
\ & =e^{\diss_{a,b} [x _{ a,b } ] }
\notag
\end{align}
Here $\Delta F _{ a,b }  \equiv F _{ b} - F _{a }  $ is the free energy change and
\begin{equation}
W _{ a,b } [ x _{ a,b } ] = \sum_{ t=a} ^{ b-1 } 
\Bigl[ E ( x _{ t+1 }, \lambda_{t+1} ) - E (x _{ t+1 }, \lambda_{t} ) \Bigr]
\end{equation}
is the  work transfered to the system during the forward process~\cite{Crooks1998, Peliti2008}. The  dissipation $\diss_{a,b} [ x _{ a,b } ] =\beta(W _{ a,b } 
[ x _{ a,b } ] - \Delta F _{ a,b })$, the irreversible increase in entropy along the forward trajectory, is proportional to the difference  between the work and the free energy change.  Note that work, free energy change and dissipation are all odd functionals under a time reversal, e.g.\ 
 $\tilde{W} _{ b,a }  [ \tilde{x} _{ b,a } ] = - W _{ a,b }  [x _{ a,b } ] $.

We can express the trajectory ensemble average of an arbitrary trajectory dependent function $\mathcal{F}[ x _{ a,b } ]$, starting from thermal equilibrium, as 
 \begin{equation}
\big\langle \mathcal{F} [ x _{a, b } ] \big\rangle _{ a,b }
\equiv 
\sum_{x _{a,b} }  P _{ a,b } [x _{ a,b } ]\
\mathcal{F} [ x _{ a,b } ]
\label{eq:pathavgforward}
\end{equation}
and similarly for the conjugate process \begin{equation}
\big\langle \tilde{\mathcal{F}} [ \tilde{x} _{b, a } ] \big\rangle _{ b,a }
\equiv 
\sum_{\tilde{x} _{b,a} }  \tilde{P} _{ b,a } [\tilde{x} _{ b,a } ]\
\tilde{\mathcal{F}} [ \tilde{x} _{ b,a } ]
\label{eq:pathavgreverse}
\end{equation}
in which $\tilde{\mathcal{F}} [ \tilde{x} _{ b,a } ] = 
\mathcal{F} [ x _{ a,b } ]  $ is an even functional under time reversal.

A key result in our development links  
two different  trajectory ensemble averages 
\begin{equation}
\Bigl\langle \mathcal{F} [ x _{a,b}] \Bigr\rangle _{ a,b }   = 
\Bigl\langle e ^{ -\diss_{0,a} [ x _{ 0,a } ]  } \ 
\mathcal{F} [x _{a,b} 
] \Bigr\rangle _{ 0,T } 
\label{eq:forwardcontract}
\end{equation}
where $0\leq a<b\leq T$.  Given a protocol $\Lambda_{0,T}$, we can extract the value of a trajectory ensemble average over a subinterval $\Lambda_{a,b}$, as if the system  began in equilibrium at an intermediate  time $a$,  by re-weighting the observations by the exponential of the dissipation from the initial to intermediate time.

This result follows directly from the work fluctuation relation [Eq.~(\ref{eq:WFT})], and the Markovian property of the dynamics.
\begin{align}
\left\langle e ^{ - \diss _{0,a} [x  _{ 0,a }  ] } \mathcal{F} [x _{a,b} ] \right\rangle _{ 0,T }
 & = 
 \left\langle e ^{ - \diss _{0,a} [x  _{ 0,a }  ] } \mathcal{F} [x _{a,b} ] \right\rangle _{ 0,b }
 \\\notag &=  
 \left\langle e ^{   - \tilde{\diss} _{b,a} [\tilde{x}  _{ b,a }  ]   }  \tilde{\mathcal{F}} [\tilde{x} _{b,a} ] \right\rangle _{ b,0 }
  \\ \notag &=  
 \left\langle e ^{   - \tilde{\diss} _{b,a} [\tilde{x}  _{ b,a }  ]   }  \tilde{\mathcal{F}} [\tilde{x} _{b,a} ] \right\rangle _{ b,a }
\\ \notag &=
\big\langle \mathcal{F} [ x _{a,b}] \big\rangle _{ a,b }   
\end{align}
We truncate the time interval of the trajectory ensemble average using the stochastic 
property in Eq.~(\ref{eq:stochastic}), apply a time reversal with the work 
fluctuation theorem in Eq.~(\ref{eq:WFT}), truncate again, and apply a second time reversal.
This result generalizes previous trajectory ensemble averages of Hummer and Szabo~\cite{Hummer2001a} and 
  Chelli et al.~\cite{Chelli2008}.

We can now use this relation to extract the thermodynamic length from far-from-equilibrium experiments. 
The discrete time analogs of the Fisher length and divergence are the cumulative Jensen-Shannon length  
 \begin{equation}
\mathcal{L}_{\text{JS}} \equiv  \sqrt{8} \sum_{ t=0 } ^{ T-1 } 
\sqrt{\JS \Bigl(\pi (x|\lambda _{t} ), \pi (x|\lambda _{ t+1 } )\Bigr) } \,
\label{eq:jslength}
\end{equation}
and cumulative Jensen-Shannon divergence~\cite{Crooks2007c},
\begin{equation}
\mathcal{J}_{\text{JS}} \equiv   8 \sum_{ t=0 } ^{ T-1 } 
\JS \Bigl(\pi (x|\lambda _{t}) , \pi (x|\lambda _{ t+1 }) \Bigr)   \ .
\end{equation}
Here, $\JS (p _{1} , p _{2} )$ is the Jensen-Shannon divergence~\cite{Lin1991,Endres2003} between two probability distributions $p _{1}  $ and $p _{2}  $,  
\begin{align}
\JS \Bigl(p _{1} , p _{2} \Bigr) 
 &  = 
 \frac{1 }{2 } \sum_x
p _{1 } (x) \ln \frac{p _{1} (x)  }{ \frac{1 }{2} [
p _{1} (x) + p _{2} (x)] } 
\\ & \ \ \
+ \frac{1 }{2} \sum_x
p _{2} (x) \ln \frac{p _{2} (x)  }{\frac{1 }{2} [
p _{1} (x) + p _{2} (x)]} \ .
\notag
\end{align}
The Jensen-Shannon length is less than the Fisher length $\mathcal{L}_{\text{JS}}\leq \mathcal{L}$, and approaches equality as the step size along the path decreases~\cite{Crooks2007c}.

We can use the contracted trajectory average Eq.~(\ref{eq:forwardcontract}), and the canonical probabilities [Eq.~(\ref{eq:canonical})] to write the 
 the Jensen-Shannon divergence between any pair of time points along the path  
\begin{align}
\label{eq:contractedjs}
\JS( \pi_t, \pi_{t+1}) 
= & \frac{1}{2}
\left\langle e^{-\mathcal{D}_{0,t}} \ln \frac{2}{1+  e^{-\mathcal{D}_{t,t+1}} }\right\rangle_{0,t+1} 
\\ &+ \frac{1}{2} \left\langle e^{- \tilde{\mathcal{D}}_{T,t+1}} \ln \frac{2}{1+   e^{-\tilde{\mathcal{D}}_{t+1,t}} }\right\rangle_{T,t} 
\notag
\end{align}
as a trajectory  average of the dissipation $\diss$ along the forward and reverse protocols.
While $\JS (\pi _{ t }, \pi _{ t+1 } ) $ is defined in terms of averages
over equilibrium probability distributions, it can be related to 
trajectory ensemble averages of processes driven arbitrary far-from-equilibrium.  
The derivation of Eq.~(\ref{eq:contractedjs}) requires the 
time reversal symmetry in Eq.~(\ref{eq:reverse}).

We now encounter an apparent complication. The dissipation $\mathcal{D}_{a,b} = \beta \left(W_{a,b} -\Delta F_{a,b} \right)$ depends on both the work and the free energy. Therefore, we must also determine the potential of mean force, the free energy as a function of $\lambda$, along the entire path. This  problem of extracting free energy profiles from out-of-equilibrium work measurements (rather than just the difference in free energy between the initial and final ensembles) has attracted recent attention~\cite{Chelli2008,Minh2008,Shirts2008}. Here, we will solve this problem by adapting  Bennett's maximum likelihood method~\cite{Bennett1976, Shirts2003,Maragakis2006, Maragakis2008b}, which, as we shall see, is intimately linked to the thermodynamic divergence~\cite{Crooks2007c}.  

Suppose we have taken measurements of the work during $N$ repetitions of a protocol $\Lambda_{a,b}$ and another $N$ measurements from the conjugate protocol $\tilde{\Lambda}_{b,a}$.  Each repetition begins in thermal equilibrium
with  the control parameter fixed at $\lambda_a$ or $\lambda_b$. 
Then the Bennett log-likelihood that the free energy change $\Delta F_{a,b}$ has a particular value is~\cite{Crooks2007c}
\begin{align}
\label{eq:bar2states}
\ell(\Delta F_{a,b}) = & \sum_{ n=1 } ^{ N  } 
\ln \frac{1 }{ 1 + e ^{ - \beta W ^{ (n) }_{a,b} + \beta \Delta 
F_{a,b} } } 
\\ &\ + \sum_{ n=1 } ^{ N  } 
\ln \frac{1 }{ 1 + e ^{ - \beta \tilde{W} ^{ (n) }_{b,a} + \beta \Delta 
F_{b,a} } } 
\notag
\end{align}
where $W^{ (n) }_{a,b}$ and  $\tilde{W} ^{ (n) } _{ b,a }  $ are the work measured between  $a$ and $b$ during the $n$th repetition of the forward and
reverse  experiment respectively.  

Next we extend this result using the contracted trajectory average [Eq.~(\ref{eq:forwardcontract})] to estimate the likelihood of the free energy change between any two points along the protocol. In particular, we can estimate the log-likelihood for the entire free energy profile by summing the log-likelihood for every pair of neighboring time points
\begin{align}
\ell\big(\{F\} \big)
= 
\sum_{ t=0 } ^{ T-1 } \Bigg[& \sum_{n=1}^{N} 
 e^{-\mathcal{D}^{(n)}_{0,t} }\ln \frac{1}{1+   e^{-\mathcal{D}^{(n)}_{t,t+1} }}
 \\ &+ \sum_{n=1}^{N} 
  e^{-\tilde{\mathcal{D}}^{(n)}_{T,t+1}} \ln \frac{1}{1+   e^{- \tilde{\mathcal{D}}^{(n)}_{t+1,t} }}
\Bigg]\ .
\notag
\end{align}
Here, $\{F\} = (F_0, F_1, \cdots, F_T)$ is the free energy profile. Since only differences in free energy are relevant, one free energy is set to zero or some other convenient reference.
Since each experimental realization of the forward and 
reverse protocols begins in 
equilibrium, the probability of 
each forward and reverse realization is $P _{ 0,N } [x _{ 0,N } ] $  and
$\tilde{P} _{ N,0 } [ \tilde{x} _{ N,0 } ] $ 
respectively.   Hence,  we have written this expression 
using the measured dissipation
$\mathcal{D}^{(n)}_{a,b}= \beta(W^{(n)}_{a,b} - \Delta F_{a,b})$ so that the relationship with Eq.~(\ref{eq:contractedjs}) is clear.   
To within an additive constant, the total Bennett log-likelihood is proportional to the cumulative Jensen-Shannon divergence. Therefore, we can simultaneously determine the potential of mean force, the thermodynamic divergence, and the thermodynamic length from the same collection of work measurements. We first  determine the free energy profile $\{ \widehat{F} \}  $  that maximizes the log-likelihood, which immediately provides an estimate of the thermodynamic divergence,
\begin{equation}
\widehat{\mathcal{J}}_{\text{JS}} = \frac{8}{N}  \left( \frac{1}{2} \ell\big(\{\widehat{F} \}\big) + T N \ln 2\right) \ .
\end{equation}
We can then calculate a maximum likelihood estimate of the thermodynamics length $\widehat{\mathcal L}_{\text{JS}}$ in Eq.~(\ref{eq:jslength}) 
using the free energy profile.

In principle, it should be possible to measure thermodynamic length in single molecule experiments. One captures a RNA hairpin in an optical tweezer, and repeatedly measure
the force on the RNA molecule as a function of extension. These far-from-equilibrium force extension
curves will then yield the potential of mean force, plus the thermodynamic length and divergence of the protocol. 
Far-from-equilibrium measurements of thermodynamic length have interesting
implications for nano-scale machines and biological motors, since 
the square of this length 
bounds the dissipation during finite time
protocols~\cite{Salamon1983a}.   
Hence, thermodynamic length is intimately connected
with the useful work that a system can perform. 
 It would, for example, be interesting to measure the thermodynamic length along the
cycle of a molecular motor. For a machine operating at a finite rate, but otherwise optimized to minimize dissipation, it is expected that the rate of change in thermodynamic length along the cycle would be constant~\cite{Salamon1985}.

\begin{acknowledgments}
We would like to thank John Chodera and David Minh  for pertinent correspondence. 
This research was supported by the U.S. Dept. of Energy, under contracts
DE-AC02-05CH11231.  E.H.F.\ thanks the Miller Institute
for Basic Research in Science for financial support.  
\end{acknowledgments}


\end{document}